# Resolved images of a protostellar outflow launched by an extended disk wind

Per Bjerkeli[1,2], Matthijs H.D. van der Wiel[1,3], Daniel Harsono[4], Jon P. Ramsey[1], and Jes K. Jørgensen[1]

[1] Centre for Star and Planet Formation, Niels Bohr Institute & Natural History Museum of Denmark, University of Copenhagen, Øster Voldgade 5–7, 1350 Copenhagen K., Denmark

[2] Department of Earth and Space Sciences, Chalmers University of Technology, Onsala Space Observatory, 43992 Onsala, Sweden

[3] ASTRON, the Netherlands Institute for Radio Astronomy, Postbus 2, 7990 AA Dwingeloo, The Netherlands

[4] Heidelberg University, Center for Astronomy, Institute of Theoretical Astrophysics, Albert-Ueberle-Straße 2, 69120 Heidelberg, Germany

**Young stars are associated with prominent outflows of molecular gas[1,2]. The ejection of gas via these outflows is believed to remove angular momentum from the protostellar system, thus permitting young stars to grow by accretion of material from the protostellar disk[2]. The underlying mechanism for outflow ejection is not yet understood[2], but is believed to be closely linked to the protostellar disk[3]. Assorted scenarios have been proposed to explain protostellar outflows; the main difference between these models is the region where acceleration of material takes place: close to the protostar itself ('X-wind'[4,5], or stellar wind[6]), in a larger region throughout the protostellar disk (disk wind[7-9]), or at the interface between[10]. Because of the limits of observational studies, outflow launching regions have so far only been probed by indirect extrapolation[11-13]. Here we report observations of carbon monoxide toward the outflow associated with the TMC1A protostellar system. These data show that gas is ejected from a region extending up to a radial distance of 25 astronomical units from the central protostar, and that angular momentum is removed from an extended region of the disk. This demonstrates that the outflowing gas is launched by an extended disk wind from a Keplerian disk. Hence, we rule out X-wind and stellar wind launching scenarios as the source of the emission on the scales we observe.**

In this work, we obtained high-angular resolution millimetre-wave observations of the region surrounding the protostar TMC1A using the Atacama Large Millimeter/submillimeter Array (ALMA). We observed the $J = 2$–$1$ rotational transition of the carbon monoxide isotopologues $^{12}CO$, $^{13}CO$, and $C^{18}O$. TMC1A is located in the Taurus Molecular Cloud (140 parsec distance), and is a protostellar system with a ~0.5 $M_{sun}$[14] protostar moving away from the solar neighbourhood at a systemic velocity of 6.4 km s$^{-1}$. It is surrounded by a circumstellar envelope with diameter of ~$10^4$ astronomical units (au)[15], a disk-like structure of radius 200 au which is inclined 55° with respect to the plane of the sky[14], and has a bipolar outflow extending at least $6 \times 10^3$ au[16] with position angle ~165° east of north. Thus far, TMC1A has been studied at spatial resolutions ranging from several thousand down to ~100 au and the disk is known to exhibit a Keplerian rotation profile[14] at radial distances ~60–100 au. The outflow, directed perpendicular to the disk, is bipolar in nature, but is most prominent on the north side of the disk[16-18].

The observations (see Methods) were taken at a spatial resolution of 6 au for TMC1A and cover the inner 200 au of the outflow as well as the disk surrounding the protostar. The $^{12}CO$ channel maps in Fig. 1 reveal the walls of the outflow cavity, while the $^{13}CO$ and $C^{18}O$ emission follows the structure of the dust continuum emission (see Extended Data Fig. 1) emanating from the 0.05 $M_{sun}$ disk[14] surrounding TMC1A. A non-disk origin for $^{12}CO$ is suggested by the significant spatial shift of the $^{12}CO$ emission relative to the $^{13}CO$, $C^{18}O$, and dust continuum emission (see Methods and Extended Data Fig. 1). The morphology of the $^{12}CO$ emission changes significantly with velocity and small-scale structure is visible in the maps (Fig. 1). These knots could represent density fluctuations in the flow[19,20], but additional observations at multiple epochs are needed to constrain their nature. The eastern cavity wall (to the left in all figures) of the blue-shifted outflow[18] is detected above the 3σ level at velocities offset by more than 2 km s$^{-1}$ from the systemic velocity, and it is clearly offset from the disk and central outflow axes indicated by the dashed lines (e.g., Fig. 1). It extends to vertical distances of more than 100 au from the disk plane, and its direction is consistent with lower resolution observations of TMC1A tracing ~1000–5000 au scales[16]. The western cavity wall of the red-shifted outflow is also detected, but at lower velocities compared to the source velocity and at a low signal-to-noise ratio. The northwestern and southeastern cavity walls are not detected, which most likely implies that the radial velocities of these two components coincide with the velocity of foreground material, and therefore suggests that the outflow is rotating (at $v_\varphi < 4$ km s$^{-1}$; see Methods). Alternatively, in order to explain their absence, the density and temperature in these regions would have to be significantly lower than in the two other cavity walls (see Methods), but this is unlikely given the intrinsic bipolar nature of protostellar outflows.

Visual inspection of the channel maps in Fig. 1 suggests that the observed outflowing gas is not launched from within a fraction of an au from the central protostar, as would be the case in an X-wind or stellar wind scenario. The corresponding Keplerian radii (plus signs in Fig. 1) are well outside 1 au for each channel map. At velocities larger than ~5 km s$^{-1}$ with respect to the systemic velocity, virtually no emission is detected along the central outflow axis. It is also clear that outflowing gas is present at large radial separations ($r \approx 50$ au) from the central star and close to the disk surface. This is not easily reconcilable with a pure X-wind scenario since the wide-angle flow streamlines predicted by such a mechanism are not in accordance with outflowing gas that has similar outflow speeds and radial separations, but a range of heights above the disk (see e.g. Fig. 2 of Shu et al. (2000; ref 21)). Consequently, the observed emission cannot be understood using a pure entrainment explanation. The channel maps also show that lower-velocity gas is present at larger distances from the central outflow axis than higher-velocity gas. This onion-like layered structure[8] is consistent with observations on larger scales[18]. We estimate the outflow launching radii[8] (footpoint radii, $r_0$) utilizing two different methods. Firstly, we fit a first order polynomial through all the pixels above the disk midplane in the $^{12}$CO channel maps, weighted by the flux density in each pixel (see Methods). This linear least-squares fit provides a direct and straightforward estimate of the footpoint radius for each velocity channel, assuming the gas travels along straight lines (see Methods). The best-fit results, presented in Fig. 2, reveal a range of footpoint radii between ~6 and ~22 au, with a trend where $r_0$ decreases with increasing velocity. Secondly, we apply steady-state magnetohydrodynamic wind theory to derive the footpoint radii[22]. Since the same launching mechanism is responsible for the transfer of both angular momentum and kinetic energy into the wind, both in the case of a disk wind and in the case of an X-wind, the outflow and rotational velocity components must be closely linked. Consequently, the footpoint radius can be determined for each position of the map (see Methods). The analysis shows the same trend as the first method, revealing footpoint radii between ~2 and ~19 au (Fig. 3). Thus, the observed emission is consistent with a scenario where a magnetic wind ejects ions from a radially extended region of the disk (which is observed to be in Keplerian rotation around a central mass of 0.4 $M_{sun}$; see Methods and Extended Data Fig. 2), that drags molecular gas along. Indeed, the inferred range of footpoint radii is consistent with a disk wind outflow mechanism, whereas, for an X-wind or stellar wind, the footpoints should be located well inside 1 au.

In the dust continuum data, the flux density distribution reveals an excess in emission relative to the underlying Gaussian profile. The strength of this feature varies slightly with azimuthal angle (most prominent on the southern side of the disk) but is located at a relatively constant radius of ~20 au (see Extended Data Fig. 3). It is at present unclear if this feature is related to the launching mechanism, but we note that the radius, interestingly, is very similar to the estimated maximum footpoint radius of the flow (Fig. 2). We interpret the observed dust emission excess as the result of a density enhancement (and perhaps an elevated dust temperature) at the edge of the outflow launching region.

We measure the specific angular momentum from the velocity field (deprojected from the line-of-sight velocity with respect to the systemic velocity) to be less than 200 au km s$^{-1}$ in the outflow and it appears to increase with distance from the protostar (Extended Data Fig. 4). This demonstrates that a significant amount of angular momentum is removed from an extended region throughout the disk. The specific angular momentum of the outflowing gas is comparable to what has previously been reported[18] for the large scale disk, i.e. 250 au km s$^{-1}$. Compared to other sources where large scale emission is observed[13,23], the value is relatively low, however. Using the values of the specific angular momentum and the outflow velocity (deprojected from the line-of-sight velocity with respect to the systemic velocity), we can define a locus in the parameter space shown in Fig. 2 of Ferreira et al. (2006; ref 13). That figure provides theoretical predictions for the relationship between these quantities, for different launching scenarios. The TMC1A outflow falls in the regime where poloidal outflow velocities are relatively low and launching radii are large. This is consistent with a disk wind launching mechanism but inconsistent with pure stellar wind or X-wind models.

Observationally, younger outflows are found to be more collimated, have smaller opening angles, and show higher gas velocities than their older counterparts[24,25]. In this regard, TMC1A does not fall into the category of the very youngest protostars, but rather in the transition period between young and old, where there is still a significant amount of material available for accretion onto the central protostar. Theoretically, X-winds naturally produce fast, well-collimated outflows[4,5], stellar winds are effective at spinning down the central protostar[6], while slow outflows and wide opening angles are most easily explained by disk wind models[7,8]. The observations presented here demonstrate that the observed TMC1A CO outflow is launched from radial distances significantly displaced from the central protostar, but, since they do not resolve the emission on scales below 6 au, we cannot exclude the possibility of an additional, confined and high-velocity component not probed by these observations. It has in fact been suggested[2,26], that a combination of different mechanisms is needed to match all of the observations, where the disk wind might be important for driving a wide-angle outflow capable of removing a large portion of the infalling envelope[24]. Generally, the most promising theories proposed for protostellar outflow ejection (X-winds, stellar winds, and disk winds) have difficulties explaining both large opening angles and bow-shaped structures simultaneously.

A well-known observational fact in meteoritics is that part of the chondritic material found throughout the Solar System has a composition consistent with having undergone thermal processing at very high temperatures only expected in the inner Solar System[27,28]. If the disk wind observed in this work extends to smaller radii where it cannot currently be resolved, it could viably form the first link in a chain that could transport thermally-processed solid material outwards in a protostellar system by allowing it to rain down on the outer part of the disk, whereas an X-wind type outflow could not[29]. The TMC1A system has an age of at most a few $10^5$ years[30]. Although these observations do not probe the very smallest scales, they show that it is possible to drive such a mechanism at times sufficiently early to couple with the formation epoch of various, chondritic components[28] in a young analogue of the Solar System.

**Acknowledgements**
The authors wish to thank Martin Bizzarro, Lars Kristensen and the anonymous referees for suggestions that helped improve the presentation and the quality of the paper. This research was supported by the Swedish Research Council (VR) through the contract 637-2013-472 to Per Bjerkeli. Matthijs H. D. van der Wiel and Jes K. Jørgensen acknowledge support by a Lundbeck Foundation Junior Group Leader Fellowship as well as the European Research Council (ERC) under the European Union's Horizon 2020 research and innovation programme (grant agreement No 646908) through ERC Consolidator Grant "S4F". Centre for Star and Planet Formation is funded by the Danish National Research Foundation. Daniel Harsono is funded by Deutsche Forschungsgemeinschaft Schwerpunktprogramm (DFG SPP 1385) The First 10 Million Years of the Solar System – a Planetary Materials Approach. The authors thank the staff at the Nordic ALMA Regional Centre node for assistance with the preparation and calibration of the data. D.H. thanks Leiden Observatory for providing the necessary adequate computing facility. This paper makes use of the following ALMA data:
ADS/JAO.ALMA#2015.1.01415.S.
ALMA is a partnership of ESO (representing its member states), NSF (USA) and NINS (Japan), together with NRC (Canada), NSC and ASIAA (Taiwan), and KASI (Republic of Korea), in cooperation with the Republic of Chile. The Joint ALMA Observatory is operated by ESO, AUI/NRAO and NAOJ.

**Author contributions**
P.B. and M.H.D.vdW. led the project and were responsible for the data reduction, analysis and writing of the observing proposal and manuscript. D.H, J.P.R and J.K.J contributed at various stages to the data reduction and analysis, discussed the results and contributed to the proposal and manuscript.

**Author information**
Reprints and permissions information is available at www.nature.com/reprints. The authors declare no competing financial interests. Correspondence and requests for materials should be addressed to P.B (per.bjerkeli@bjerkeli.se).

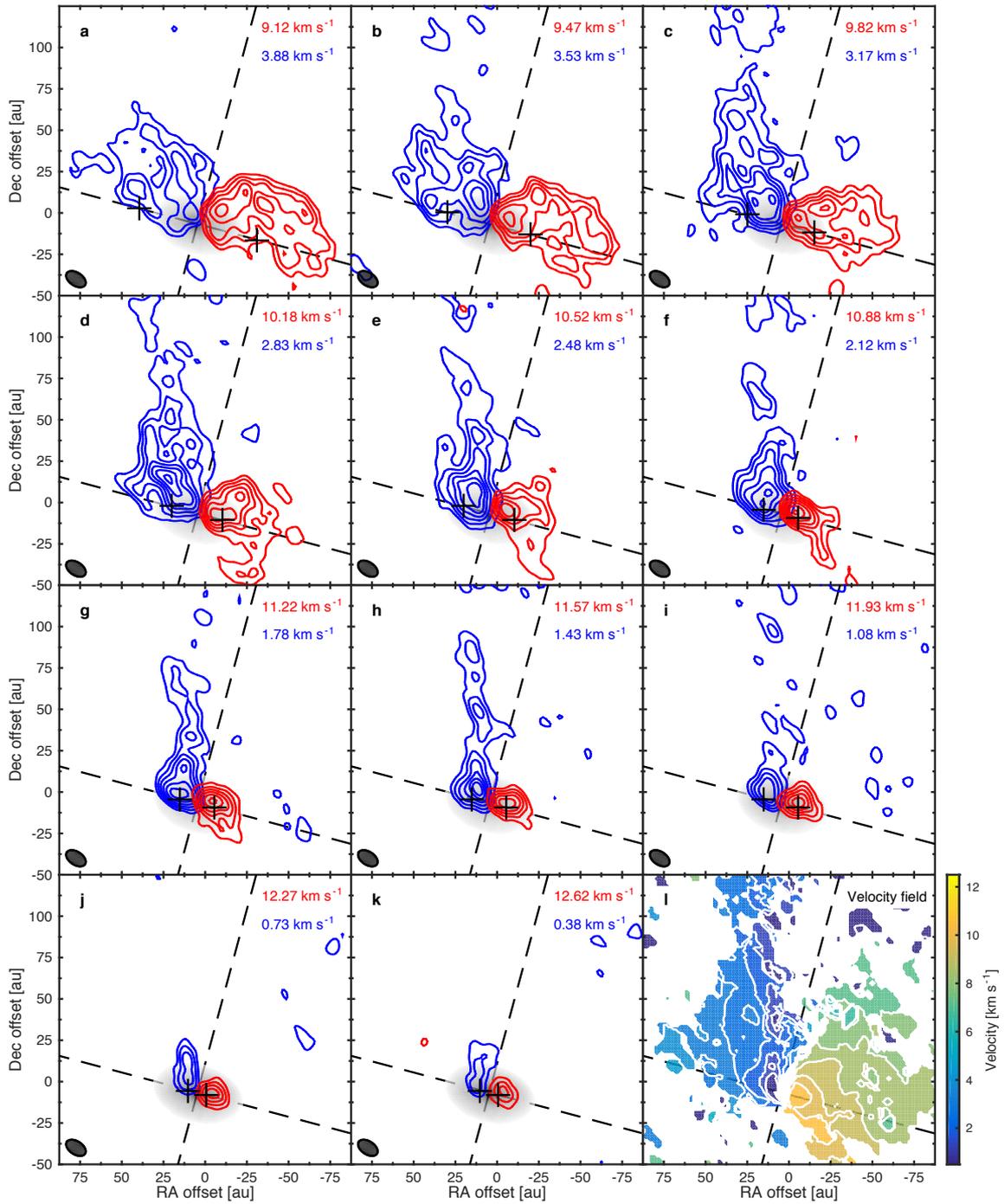

**Figure 1. $^{12}$CO channel map of the region.** Contours are from $3\sigma$ in steps of $1.5\sigma$ ($\sigma$ = 0.8 mJy/beam). Blue and red contours represent the emission that is blue-shifted and red-shifted, respectively, with respect to the systemic velocity (6.4 km s$^{-1}$). The central channel velocities, the synthesized beam, and the radii for the corresponding Keplerian velocities (plus signs) are indicated in each panel. Dashed lines indicate the directions of the disk and the perpendicular outflow axis. The dust continuum emission is shown in greyscale. The lower right panel shows the velocity field of the $^{12}$CO emission (moment 1 map).

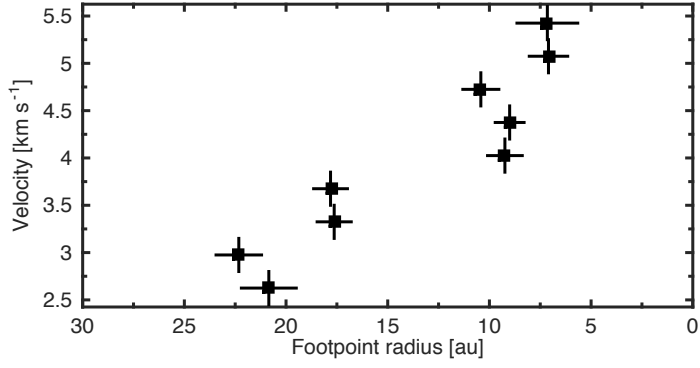

**Figure 2. Direct measure of the disk wind launching point.** All pixels above the disk midplane, where the emission flux density is above 3σ, are fit using a first order polynomial. The launching point (footpoint radius, $r_0$) is where the best-fit line crosses the disk midplane. Velocity is the absolute value of the observed line-of-sight velocity with respect to the systemic velocity of the protostar (6.4 km s$^{-1}$). Black lines represent the 2σ confidence interval on the fit and the velocity resolution of the observations. A trend, where higher velocity gas corresponds to a smaller estimated footpoint radius, is visible.

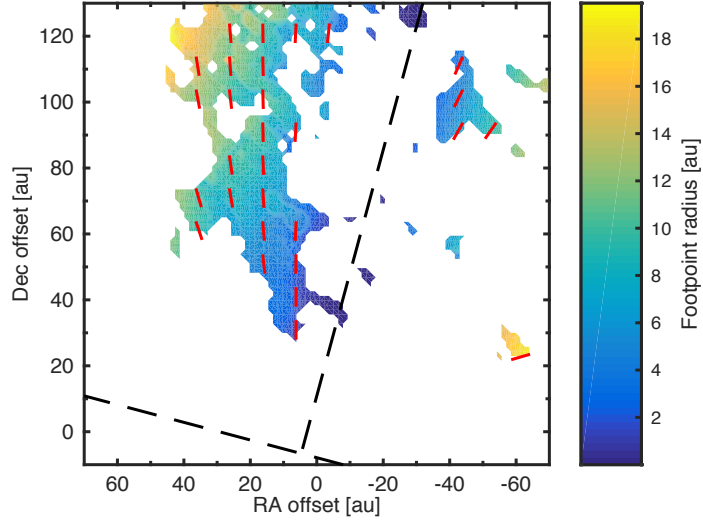

**Figure 3: Disk wind launching point, applying steady-state magnetohydrodynamic wind theory.** The footpoint radius ($r_0$) is derived under the assumption that the same launching mechanism is responsible both for the angular momentum and kinetic energy extraction into the flow[22]. All pixels in which the outflow velocity is smaller than the local escape velocity have been masked out. The central outflow axis and the disk midplane are indicated with dashed lines. For a selection of pixels, the red lines point to the derived launching point/footpoint; the length of the line is arbitrary.

## Methods

**ALMA observations and data processing**

TMC1A was observed with ALMA on 2015 October 23 and 30. The observations presented here are part of the Cycle 3 program 2015.1.01415.S. The primary beam of the ALMA 12 m dishes covers a field of 22 arcseconds in diameter around TMC1A and the source was observed in the $J = 2–1$ rotational transitions of $^{12}CO$, $^{13}CO$ and $C^{18}O$ at 230.5, 220.4, and 219.6 GHz, respectively. 49 antennas in the 12-m array were used during the observations providing baselines in the range 85–16196 m (=65–12450 k$\lambda$). The CO observations were carried out at a spectral resolution of 244 kHz (≈0.32 km s$^{-1}$) and the total bandwidth is 117 MHz. In addition to these basebands, one baseband was used for continuum observations where the spectral resolution was set to 976 kHz (≈1.35 km s$^{-1}$) and the total bandwidth was 1875 MHz. The precipitable water vapour during the observations was between 0.28 and 0.63 mm. The phase centre of the observations was $\alpha_{2000}$ = 04h39m35.2s, $\delta_{2000}$ = +25°41′44.27′′. The peak of the continuum emission is slightly offset from this coordinate, i.e. by +0.03′′,−0.05′′ (+5,−7 au corresponding to $\alpha_{2000}$ = 04:39:35.2, $\delta_{2000}$ = +25°41′44.23′′). The X-ray source J0440+2728 was used as phase calibrator and the blazar J0510+1800 was used as bandpass calibrator. The flux calibration uncertainty is estimated to be less than 10%.

The data calibration and imaging was carried out in CASA[31] (version 4.5.0) and followed standard procedure. The continuum is subtracted in the uv-domain by fitting a constant to the line free channels. The calibrated visibilities for the continuum are transformed into the image domain using the CLEAN algorithm[32] with Briggs weighting and the robust parameter set to 0.5. For the line emission, natural weighting is used in order to provide the highest signal-to-noise ratio. To improve the signal-to-noise ratio, we used a uv taper at 0.04 arcseconds for the continuum and $^{12}CO$ maps (shown in Fig. 1) and a uv taper at 0.10 arcseconds for the $^{13}CO$ and $C^{18}O$ maps (Extended Data Fig. 1). All spectral line output images have a spectral resolution of 0.35 km s$^{-1}$. The interferometric nature of the observations leads to spatial filtering of large-scale structures, which, for these observations, leads to a maximum recoverable scale of 0.4 arcseconds (~60 au at 140 pc distance). This implies that we do not detect any emission that is extended over scales larger than 60 au and we do not detect emission at velocities below 2 km s$^{-1}$ relative to the systemic velocity of 6.4 km s$^{-1}$. Hence, we probe material moving with a velocity significantly offset from any extended emission in the system[18] and we do not recover any foreground emission from the envelope.

**Analysis of the continuum and spectral line maps**

Each velocity channel is analysed individually using MATLAB. For the presented maps, the first contour is always at 3$\sigma$. The rms level of each map is calculated in a 1.3 by 1.3 arcsecond emission free region located at 1.5 arcsecond distance from the continuum peak. The presented data has not been corrected for the primary beam response. This has no effect on the presented maps, since the correction is less than 1% within 2 arcseconds of the phase centre of the observations.

*Origin of the emission*

A crucial part of the analysis is to identify the molecular emission that arises from the disk. This can be done through direct comparison of the $^{12}CO$, $^{13}CO$ and $C^{18}O$ emission. The line ratios between the isotopologues are close to one, suggesting that the medium is optically thick in $^{12}CO$. To estimate the optical depths, the emission of the isotopologues at $v \approx 9$ km s$^{-1}$ is used. Assuming a kinetic temperature of 100 K and adopting isotopic ratios of 60 and 550 for $^{13}CO$ and $C^{18}O$, we calculate the optical depths ($\tau$) to be 0.04, 0.2 and 25 for $C^{18}O$, $^{13}CO$ and $^{12}CO$, respectively. Even in the line-wings, the optical depth of $^{12}CO$ is much greater than 10. The spatial distribution of $^{12}CO$ differs significantly from that of $^{13}CO$ and $C^{18}O$; the latter two roughly trace the rotation of the disk (Extended Data Fig. 1). Furthermore, $^{12}CO$ is not detected in the outer parts of the disk, while $^{13}CO$ and $C^{18}O$ are. This implies that the $^{12}CO$ in the disk is invisible. In Extended Data Fig. 1, the extent of the disk is derived from integrating the line wings, avoiding the line centre at $v < 2$ km s$^{-1}$. If the $^{12}CO$ emission indeed arises from the disk, the only way to explain the difference in isotopolog distribution is to have foreground material that blocks out the $^{12}CO$ emission that lies at > 2 km s$^{-1}$ from the systemic velocity, which is unlikely: Observations[33] and modelling[34] of the ambient cloud indicate a median FWHM of 1.2 km s$^{-1}$. It is thus not entirely clear why no $^{12}CO$ emission is detected in the disk. However, at large radii, the Keplerian speed approaches 2 km s$^{-1}$, and thus the emission could be absorbed by the foreground ambient material. A simple power-law disk model is used to estimate the CO isotopologue emission arising from a Keplerian disk. The continuum radiative transfer tool RADMC-3D[35] and non-LTE molecular excitation analysis[36] are used to simulate the predicted molecular emission. The models were processed through CASA using the sm ("simulation") tool in order to simulate the visibilities given the antenna positions. The models indicate that the observed spatial distribution of $^{12}CO$, $^{13}CO$ and $C^{18}O$ should be cospatial in the case of a pure Keplerian disk.

To examine the rotation of the disk on the scales where the outflow is launched, we also fit 2D Gaussian distributions to all individual channel maps for $^{13}$CO and C$^{18}$O in order to find the peak positions. We exclude the $^{12}$CO emission from this analysis due to the significant contribution from the outflow. The analysis allows us to conclude that the $^{13}$CO and C$^{18}$O emission close to the launching region is Keplerian in nature down to radial distances of ~20 au from the protostar. The central mass is estimated at 0.4±0.1 $M_{sun}$ (Extended Data Fig. 2) taking the inclination angle into account ($i = 55±10°$), i.e., slightly lower than previous estimates[14,18], and closer to what is obtained when modelling the emission with a rotating infalling envelope[37].

The offset between the northeastern cavity wall and the central flow axis, combined with the non-detection of emission from the northwestern and southeastern cavity walls, suggests that the outflow is rotating. The radial velocity of the absorbed outflow components would fall close to the systemic velocity in the case where the ratio between the outflow and rotation velocity components ($v_{out}$ and $v_{\varphi}$, respectively) is close to tan($i$). In the case where $^{12}$CO predominantly traces outflowing gas, this naturally also explains why we see redshifted emission close to the disk surface, since this is the region where the trajectory of the gas has not yet reached its asymptotic direction. In the analysis presented in this Letter, we calculate the rotation and the outflow velocity components from the observed line-of-sight velocities, corrected for the systemic velocity and deprojected by the inclination angle of the outflow (assumed to be perpendicular to the disk), i.e. $v_{\varphi} = v_{los}/\sin(i)/2$ and $v_{out} = v_{los}/\cos(i)/2$. We thus assume that the outflow is symmetric and that the rotational velocity is constant along the outflow at the scales that we observe. Rotation is also hinted at in the moment 1 map presented in Fig. 2 of Aso et al. (2015; ref 18). Although these observations probe the gas on larger scales (where most gas in the northern outflow lobe is blueshifted), it is obvious that velocities are redshifted with respect to the observer in the northwestern cavity wall and at small distances from the protostar.

*The launching radius*
The launching radius (footpoint radius, $r_0$) of the outflow is determined using a first order polynomial fitting ($z = A(r - r_0)$; $z$ is the distance above the disk midplane, $r$ is the distance from the central outflow axis and $r_0$ and $A$ are free parameters) of flux-weighted positions for each channel, and in each pixel where the signal-to-noise ratio is larger than 3 (Fig. 2). Since we are primarily interested in the outflowing gas traced by $^{12}$CO, we exclude the following regions of parameter space. Firstly, we only consider gas at a velocity of more than 2 km s$^{-1}$ offset from the systemic velocity. This is the Keplerian velocity of the disk at 100 au, and any envelope emission on larger scales will have a velocity lower than this value. Secondly, we exclude velocities > 5.5 km s$^{-1}$ with respect to the systemic velocity, since the outflow emission at these higher velocities is confined to within 20 au of the disk midplane. This analysis does not take into account that, at any given velocity, the launching region can be extended. However, it provides a straightforward and direct estimate of where the outflow is launched. We consider only the gas located to the east of the central outflow axis, since this is the only region where we attain a sufficient signal-to-noise ratio to perform a quantitative analysis. This is also the cavity wall where the emission is most extended. The slope of the line and the crossing point (footpoint radius) of the disk midplane, are thus determined by the flux density distribution across the map. We include all Nyquist sampled data points on the northern side of the disk outlined by the continuum. However, the velocity observed close to the disk surface can be significantly lower than the local escape velocity and we therefore also performed the analysis excluding all data points within 20 au of the disk surface (deprojected distance). The resulting values for $r_0$, however, are indifferent to the choice of cut-off height, and our conclusions are therefore not affected. We fit the outflowing gas by a first order polynomial to avoid making too many assumptions on the geometrical structure of the flow. We do acknowledge that, theoretically, the gas will not follow straight lines in the immediate vicinity of the launching region. To test the robustness of our results, we also fit a second order polynomial to the emission ($z = A(r - r_0)^2$). The derived footpoint radii are similar and we conclude that the choice of exponent does not affect our scientific conclusions. If we instead assume an intercept of zero during the fitting procedure (i.e. $z = Ar^2$; ref 38), the goodness of fit decreases because the observed geometry stands significantly more steeply than can be modelled by, in fact, any simple polynomial with a zero intercept (e.g. $z = Ar^B$).

In an independent analysis, we estimate the footpoint radius for each detected pixel in the $^{12}$CO map, using Eq. 4 of Anderson et al. (2003; ref 22). The characteristic velocity in each position is taken from the computed velocity field (Fig. 1) and in each position, the velocity is decomposed into two components accounting for inclination and the systemic velocity: the rotational velocity and the outflow velocity. It should therefore, be noted that the estimated footpoint radius can be affected by entrained gas and/or asymmetries in the flow, since this increases the uncertainty on the magnitude of the velocity components. Further, such an analysis can only be carried out in the ballistic regime, and for that reason, we mask out all pixels where the local escape velocity exceeds the outflow velocity. The estimated footpoint radius for each position is presented in Fig. 3. An illustration of where the outflow is launched is shown in Extended Data Fig. 5. In both figures, straight lines point towards the launching point.

*Dust continuum emission*

To examine the continuum emission from the disk, we fit a Gaussian profile to the emission as a function of the radius, deprojected by the inclination angle of the system, for all azimuthal directions. This reveals an enhancement in the emission around 20 au at the 1 mJy/beam level (see Extended Data Fig. 3), which is consistent with the estimated footpoint radius for the lowest velocity channels. Since the emission cannot easily be explained by an analytical function, we exclude all data points from the Gaussian fit where the enhancement is most prominent, i.e., between 12 and 33 au. The distance to the peak position of this enhancement does not vary significantly with azimuthal angle (see Extended Data Fig. 3).

*Angular momentum of the outflowing gas*

To estimate the specific angular momentum of the outflowing gas, we use the $^{12}$CO velocity field. The specific angular momentum is calculated as the product of the rotational velocity, and the distance to the central axis of the blue-shifted outflow (see Extended Data Fig. 4). The uncertainty on the rotation velocity is dominated by the uncertainty on the inclination angle (~10 degrees), since the uncertainty in observed velocity is negligible in comparison.

**Code availability**

The code RADMC-3D, used for the Keplerian disk modelling, is available at: http://www.ita.uni-heidelberg.de/~dullemond/software/radmc-3d/. We have opted not to make the non-LTE molecular excitation code available due to the lack of documentation and the non-trivial nature of its usage.

**Data availability**

The datasets generated and/or analysed during the current study will be available in the ALMA archive, https://almascience.nrao.edu/alma-data/archive, 10 December 2016 and are also available from the corresponding author upon reasonable request.

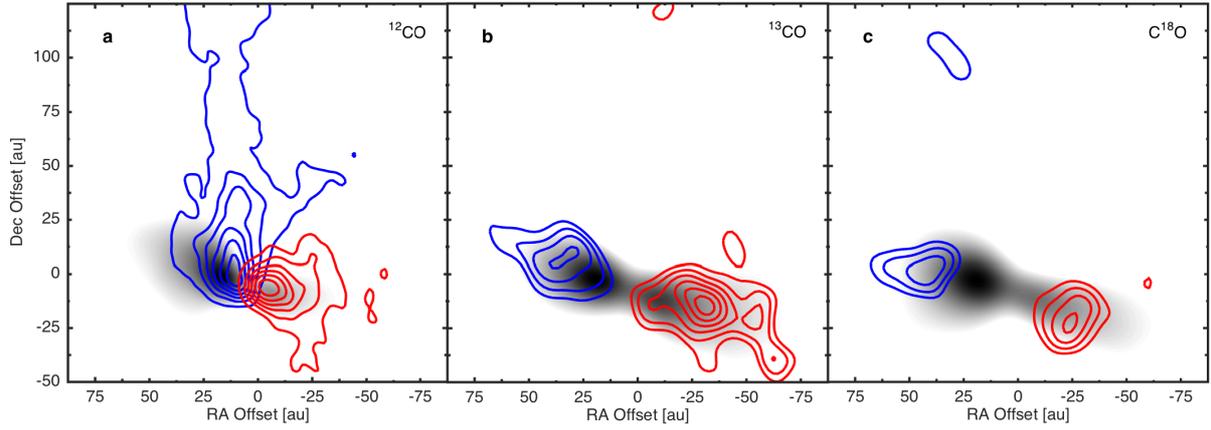

**Extended Data Figure 1. Comparison between integrated emission for $^{12}$CO, $^{13}$CO, and C$^{18}$O.** Contours are from $3\sigma$ in steps of $3\sigma$ for $^{12}$CO (a) and $1\sigma$ for $^{13}$CO (b) and C$^{18}$O (c). $\sigma$ = 4 mJy/beam for $^{12}$CO and $\sigma$ = 5 mJy/beam for $^{13}$CO and C$^{18}$O. Red-shifted and blue-shifted emission is integrated from 2.5 to 10 km s$^{-1}$ with respect to the systemic velocity. The corresponding integrated emission from the power-law disk model is shown in greyscale.

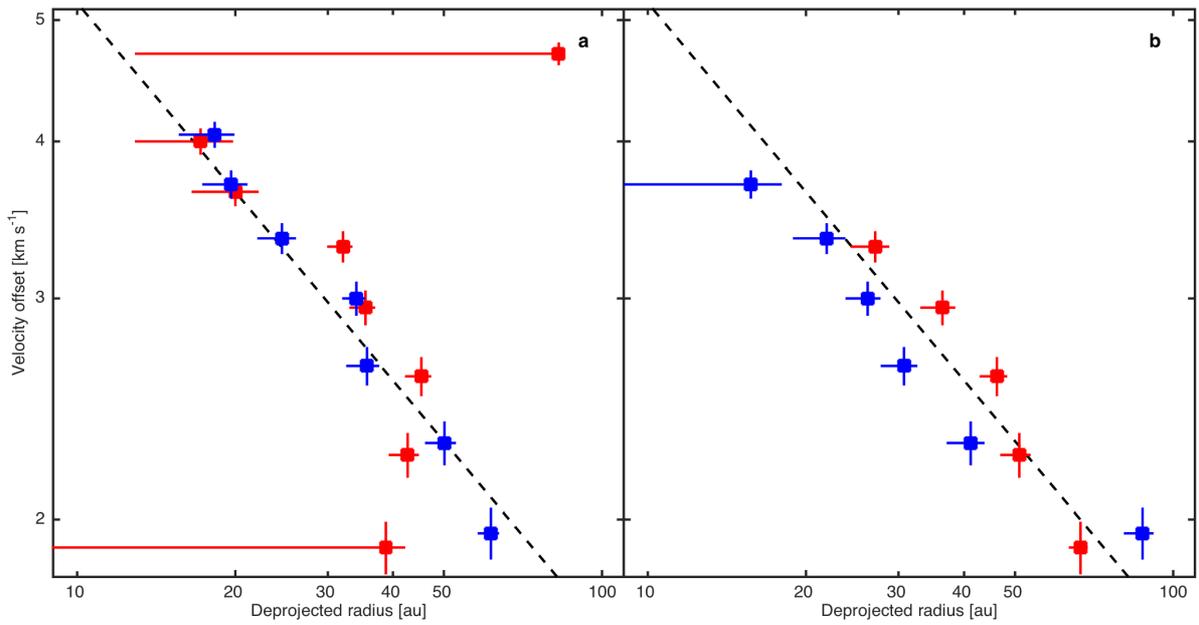

**Extended Data Figure 2. Position-velocity diagram for $^{13}$CO and C$^{18}$O.** Velocity of a: $^{13}$CO, and b: C$^{18}$O versus position, using an inclination angle of 55°. The dashed curve is indicative of Keplerian rotation around a 0.4 $M_{sun}$ star. The red and blue colours indicate the red- and blue-shifted components, respectively. Error bars show the standard deviations of the Gaussian fits in position and the velocity resolution.

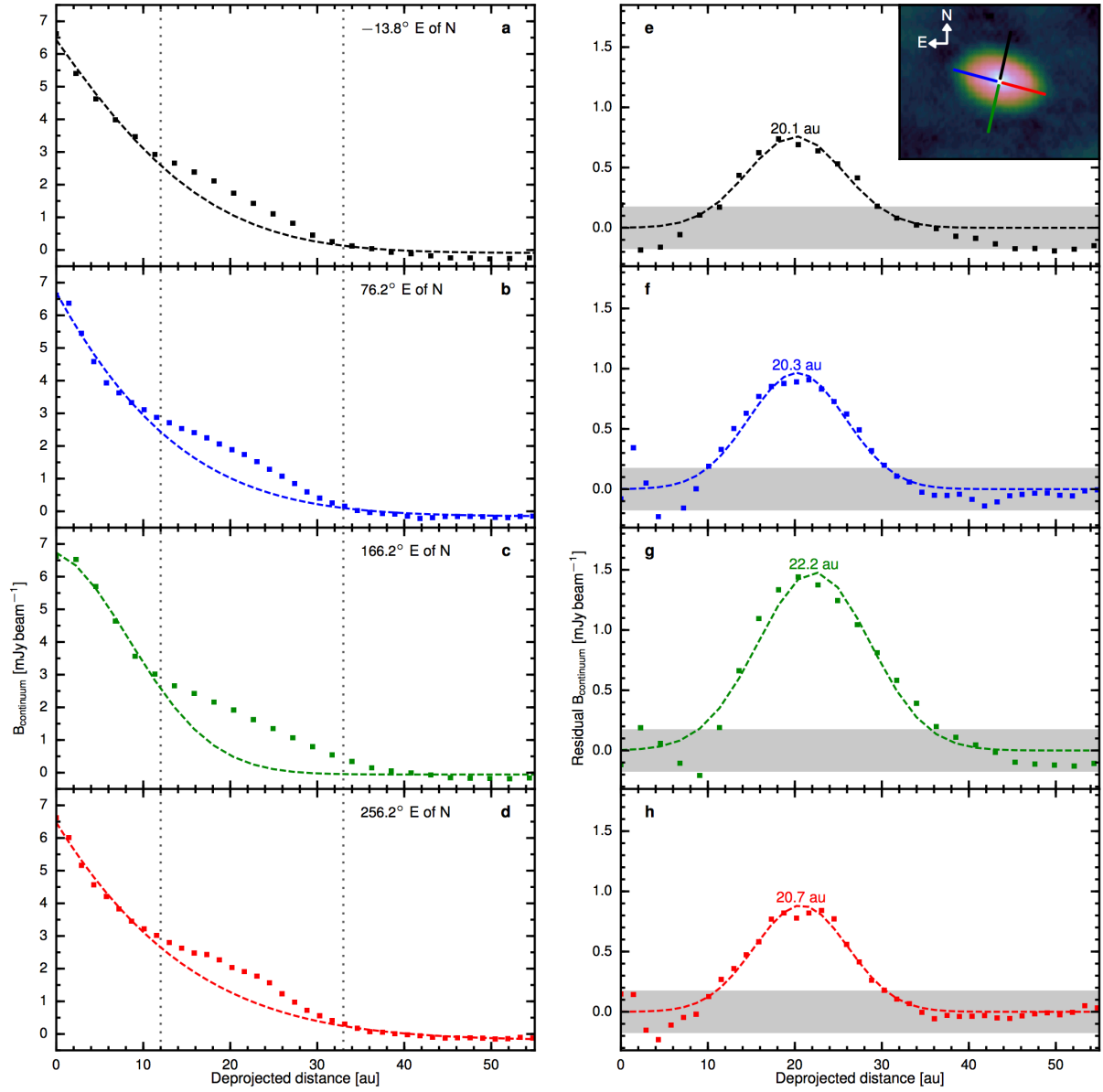

**Extended Data Figure 3. Enhancement in dust continuum emission.** *a - d:* observed radial continuum brightness profile (square points) at the four position angles (PA) indicated in the inset in the top right. PA=76° corresponds to the long axis of the disk on the northeastern side where the blue-shifted northern outflow is launched. A Gaussian fit is overlaid as a dashed line. *e - h:* residual intensity after subtracting the fits shown in the left column (square points), and a Gaussian fit (dashed line) to determine the peak location of the enhancement. The grey filled area denotes the 2σ root-mean-square noise in the continuum map.

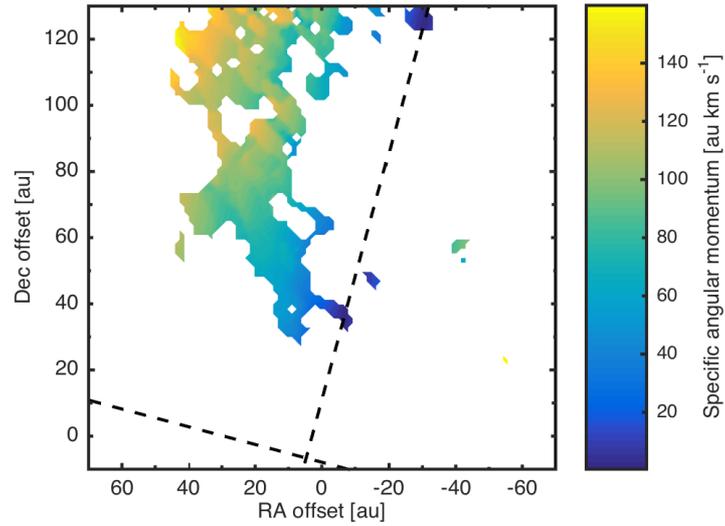

**Extended Data Figure 4. Specific angular momentum derived from the velocity field.** The colour map shows the specific angular momentum and black dashed lines show the position angle of the outflow and the disk.

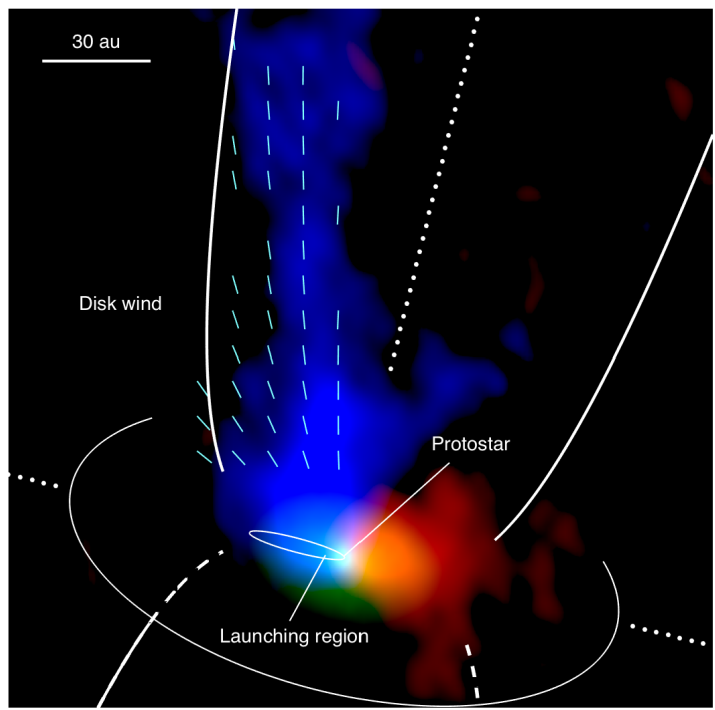

**Extended Data Figure 5. Inferred launching region of the disk wind.** The illustrative figure is overlaid on a three colour background image, showing the blue-shifted (blue) and red-shifted $^{12}$CO emission (red) together with the continuum emission (green). The outflow emission is integrated from ± (2.5 − 10) km s$^{-1}$ with respect to the systemic velocity 6.4 km s$^{-1}$. The outline of the disk and the outflow and the disk and outflow axes are indicated with white lines. Dashed lines are the same as in Fig. 3.